\documentclass[twocolumn,english]{revtex4}
\usepackage[T1]{fontenc}
\usepackage[latin1]{inputenc}
\usepackage{amsmath}
\usepackage{graphicx}

\makeatletter
\usepackage{babel}
\makeatother
\begin{document}

\title{QLib - A Matlab Package for Quantum Information Theory Calculations
with Applications}

\author{S. Machnes}

\begin{abstract}
\textbf{Developing intuition about quantum information theory problems
is difficult, as is verifying or ruling-out of hypothesis. We present
a Matlab package intended to provide the QIT community with a new
and powerful tool-set for quantum information theory calculations.
The package covers most of the {}``QI textbook'' and includes novel
parametrization of quantum objects and a robust optimization mechanism.
New ways of re-examining well-known results is demonstrated. QLib
is designed to be further developed and enhanced by the community
and is available for download at} \texttt{\textbf{www.qlib.info}}
\end{abstract}

\address{School of Physics and Astronomy, Beverly and Raymond Sackler Faculty
of Exact Sciences, Tel-Aviv University, Tel-Aviv 69978, Israel}

\date{July 31, 2007}

\maketitle

\section{Motivation}

Advances in theory are often fore-shadowed by intuition. But the mathematical
structures governing multipartite and even bipartite states and unitary
transformations are complex, which makes many problems difficult to
explore and intuition hard to develop. Analyzing problems analytically
is often a time-consuming process. Validation of hypothesis is laborious
and searching for counter-examples is a lengthy endeavor. The QIT
community will probably benefit from tools to accelerate these processes.

The use of computers for theoretical mathematics is a well established,
with a specialized journal \protect\cite{Journal of Experimental Mathematics},
textbooks \protect\cite{book-1,book-2,book-3} and numerous papers. Wolfram
research defines \textit{experimental mathematics} as ''a type of
mathematical investigation in which computation is used to investigate
mathematical structures and identify their fundamental properties
and patterns'' \protect\cite{wolfram}. Bailey and Borwein\protect\cite{B and B}
use the term to mean the methodology of doing mathematics that includes
the use of computation for

\begin{itemize}
\item Gaining insight and intuition
\item Discovering new patterns and relationships
\item Using graphical displays to suggest underlying mathematical principles
\item Testing and especially falsifying conjectures
\item Exploring a possible result to see if it is worth formal proof
\item Suggesting approaches for a formal proof
\item Replacing lengthy hand derivations with computer-based derivations
\item Confirming analytically derived results
\end{itemize}
As the benefits of tools such as Mathematica, Matlab and Maple are
clear, there is strong indication that field-specific software for
experimental theoretical quantum information would be advantageous.
QLib \protect\cite{QLib} is an attempt to provide such a tool.

\section{Overview of Capabilities}

QLib provides the tools to manipulate density matrices, separable
states, pure states, classical probability distributions (CPDs) as
well as unitary and Hermitian matrices. All of which are supported
with any number of particles, and any number of degrees of freedom
per particle. The following functions are provided to manipulate these
objects:

\begin{itemize}
\item \textbf{Entanglement calculations}: pure state entanglement, concurrence,
negativity, tangle, logarithmic negativity, entanglement of formation,
relative entanglement, robustness, PT-test (Peres Horodecki), Schmidt
decomposition and singlet fraction.
\item \textbf{Entropy}: Shannon, Von Neumann, linear entropy, relative entropy,
participation ratio, purity
\item \textbf{Measurements}: Orthogonal (to multiple collapsed states or
to a single mixture) , POVM, weak measurements
\item \textbf{Object transformation:} 

\begin{itemize}
\item Reorder particles, partial trace, partial transpose
\item Transform objects to/from the regular representation to a tensoric
representation with one index per particle if the original object
was a vector, or two indexes per particle if the object was a matrix
\item Convert to/from computational base to the base of SU(n) generators
\end{itemize}
\item \textbf{Distance measures}: Hilbert-Schmidt, trace distance, fidelity,
Kullback-Leibler, Bures distance, Bures Angle, Fubini-Study
\item \textbf{Miscellaneous}: Majorization, mutual information, spins in
3D, famous states, famous gates
\end{itemize}
QLib provides \textbf{parametrizations for all objects of interest},
density matrices, separable states, pure states, CPDs, Hermitian matrices
and unitary matrices. In other words, these object are representable
as points in a parameter space. This allows, for example, to generate
random separable states or random unitary matrices. For details of
each parametrization and its theoretical background, please refer
to the on-line help. As an example, details of two unitary-matrix
parametrizations and of one separable density matrix parametrization
are presented in \ref{sec:Sample-Applications}. \medskip{}
The \textbf{robust optimization capabilities} provided with QLib,
allows searching for extrema of functions defined over these spaces.
The optimization is performed by alternating stages of hill-climbing
and simulated annealing while applying consistency requirements to
the output of the stages. Current experience with the optimization
feature suggests that the search succeeds in locating the global extrema
in a surprising majority of the cases %
\footnote{Note that care should of always be taken when dealing with the results
of numeric optimization%
}. \medskip{}

Finally QLib provides a wide selection of general purpose utilities
which, while are not quantum-information specific, go a long way towards
making the use of QLib productive and simple:

\begin{itemize}
\item Linear algebra: Gram Schmidt, spanning a matrix using base matrices,
checking for linear independence, etc.
\item Numerics: Approximately compare, heuristically clean-up computation
results from tiny non-integer and/or tiny real/imaginary parts, etc.
\item Graphics: Quickly plot out functions in 2 and 3d, smoothing and interpolation
techniques for noisy or sparse data, etc.
\end{itemize}

\section{Getting started }

QLib, available at \texttt{\textbf{www.qlib.info}}, has been designed
for easy use. An \textit{Installation Guide} and a \textit{Getting
Started Guide} are available on the website, and over a dozen demos
are provided as part of QLib, to help you get started.

In addition, on-line help is available: simply type \texttt{\textbf{help
qlib}} at the Matlab prompt for an overview of functionality or get
function-specific help, e.g. \texttt{\textbf{help partial\_trace.}}

Finally, user forums are available to ask questions and discuss QLib
issues, and forms are provided to request new features or report bugs.

\smallskip{}

\section{\label{sec:Sample-Applications}Sample Applications}

Following are a number of QLib usage example which were selected both
for their ability to demonstrate QLib capabilities and for their relatively
simple structure and simple theoretical background, so that they may
be quickly understood by a wide range of readers.

\subsection{Entanglement of Superpositions}

Recently, some work has been done regarding the entanglement of superpositions
\begin{equation}
|\Gamma\rangle=\alpha|\Psi\rangle+\beta|\Phi\rangle\end{equation}

An upper limit to the entanglement of $|\Gamma\rangle$ has been proposed
by Linden, Popescu and Smolin \protect\cite{ent of superpos - LPS}. Further
work by Gour \protect\cite{ent of superpos - Gour} added a tighter upper
bound and a lower bound. Unfortunately, the analytical form of these
bounds make it difficult to get a good intuitive feel as to whether
they are relatively tight or whether there is still significant room
for improvement. QLib provides us with convenient tools with which
to explore the problem. See figure 1.

\begin{figure}
\includegraphics[scale=0.5]{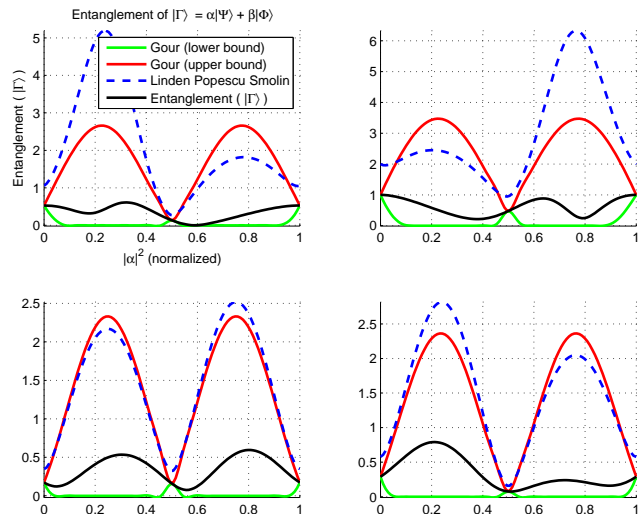}

\caption{(color on-line) Limits on entanglement of superpositions of randomly
selected states. Gour appears to match the goal function for at both
ends and when $|\alpha|=|\beta|$. Generally speaking, Gour is indeed
tighter than LPS, but not always (note the top-left sub-figure at
$|\alpha|^{2}=0.55)$. Also of note is the complex behavior exhibited
by the entanglement of superposition and the relative un-tightness
of existing bounds.}
\end{figure}

To create the graphs above, QLib's basic primitives have been used
(computation of entanglement for a pure state, normalization of a
pure state, etc), as was the capability to generate random pure states.
Finally the optimization capabilities are also put to use, as Gour's
bounds are defined in terms of maximizing a function over a single
degree of freedom for given $\Psi$,$\Phi$, $\alpha$ and $\beta$,
which requires that every point along the Gour limit lines above be
computed by an optimization process.

\subsection{Maximally Entangled Mixed States}

Over the years there has been keen interest in the question of MEMS,
Maximally Entangled Mixed States, which cannot be made more entangled
(as measured by some measure) with any global unitary transformation
\protect\cite{MEMS - Verstraete,MEMS - Ishizaka,MEMS - Wei}. QLib can assist
in exploration of this problem by searching for the most-entangling
unitary transformation. Parametrization of unitary transformations
is done either by generalized Euler angles \protect\cite{SU(n) param} or
with the more naive\begin{equation}
U=e^{i\sum_{k=1}^{n^{2}}\theta_{i}\mathbf{g}_{i}}\end{equation}

with $\mathbf{g}_{i}$ being the $U(n)\mbox{ generators.}$

In this particular example, we have explored the maximal entanglement
possible for the separable diagonal density matrix\begin{equation}
\left(\begin{array}{cc}
p & 0\\
0 & 1-p\end{array}\right)\otimes\left(\begin{array}{cc}
q & 0\\
0 & 1-q\end{array}\right)\label{eq:MEMS}\end{equation}

QLib can help discover the dependency of the maximal entanglement
on $p$, $q$ by locating the MEMS associated with the initial density
matrix and visualizing various options for $p$, $q$ dependence.
See figure 2. 

\begin{figure}
\includegraphics[scale=0.55]{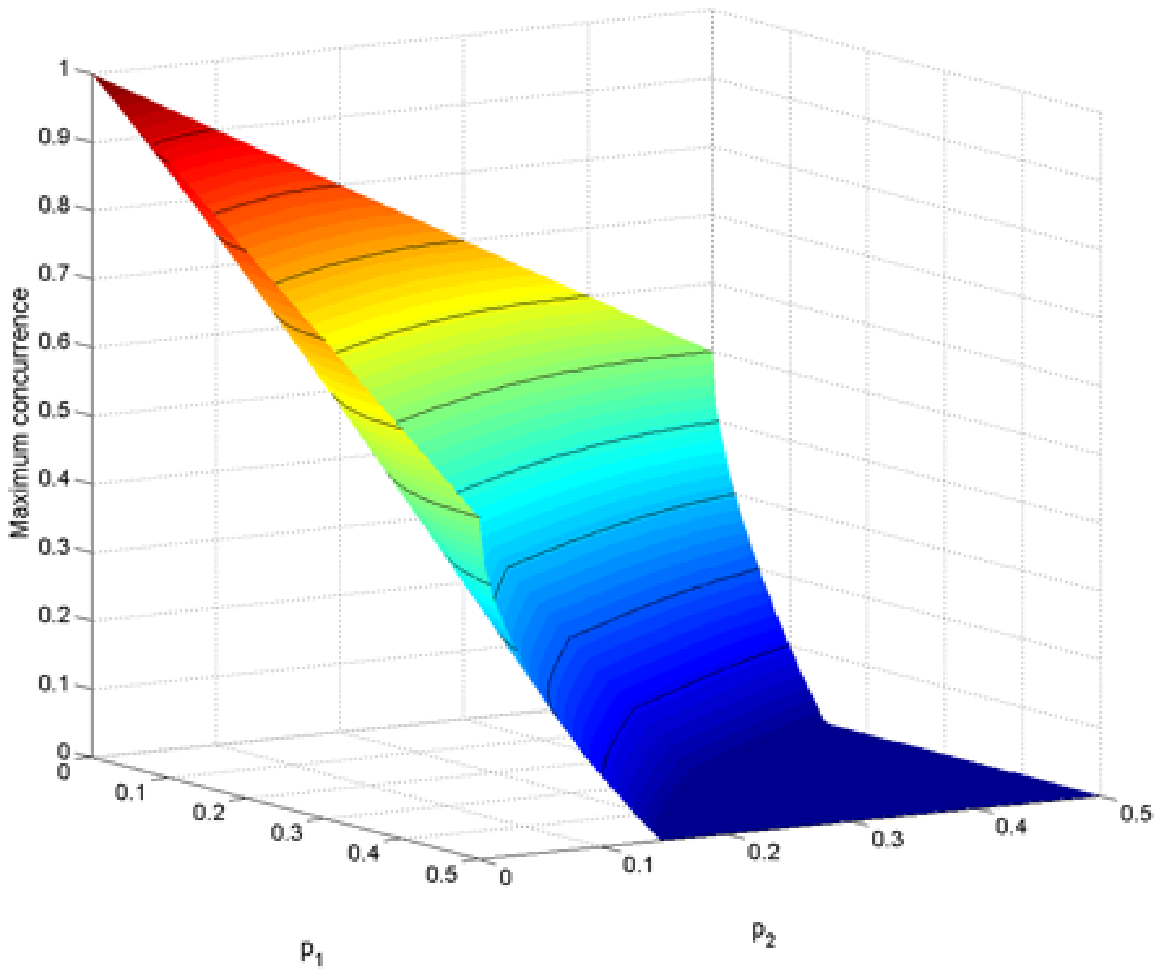}\smallskip{}
\includegraphics[scale=0.5]{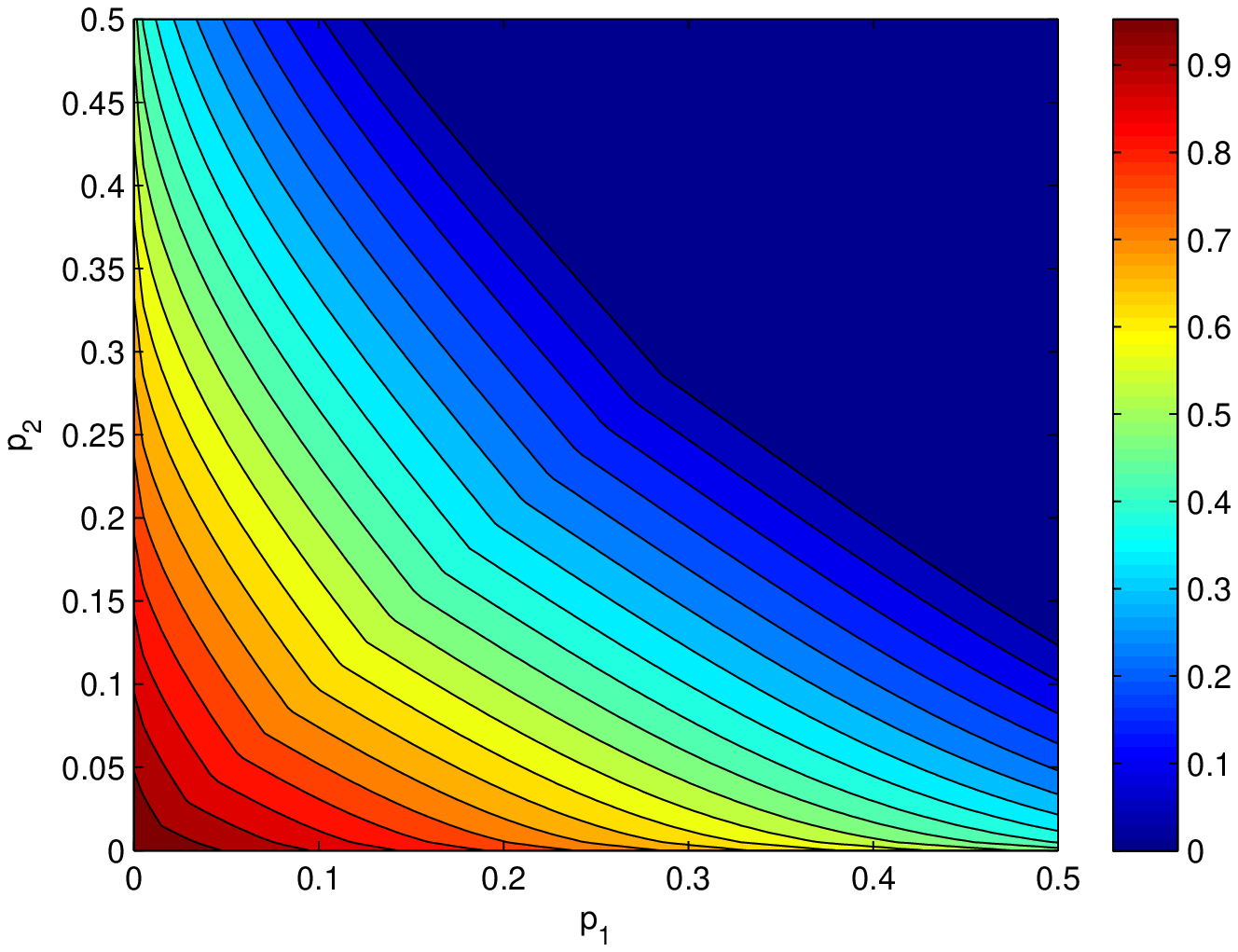}

\caption{(color on-line) Maximal concurrence for eq. \ref{eq:MEMS}. The $(p,q)=[0..0.5,0..0.5]$
space was explored with a resolution of $0.005$, for a total of $10,000$
points, for each of which an optimization of the concurrence over
the space of SU(4) unitaries has been performed. The maximal concurrence
is shown both as a function of $p$ and $q$ (in 3d, above), and as
a contour plot (below) showing the dependence of the maximal concurrence
on the trace distance between the single-particle initial density
matrix, $\left(\protect\begin{array}{cc}
p & 0\protect\\
0 & 1-p\protect\end{array}\right)$, and the fully mixed state for a single qubit $\left(\protect\begin{array}{cc}
\frac{1}{2} & 0\protect\\
0 & \frac{1}{2}\protect\end{array}\right)$.}

\end{figure}

\subsection{Bloch {}``Hyper-sphere''}

It is well known that a single qubit may be represented using the
$U(2)$ generators as\begin{equation}
\rho_{1\, qubit}(\overrightarrow{n})=\frac{1}{2}\left(\mathbf{1+}\overrightarrow{n}\cdot\overrightarrow{\mathbf{\sigma}}\right)\end{equation}

with a pure state iff $\left\Vert \overrightarrow{n}\right\Vert =1$.
This suggests a trivial generalization to higher dimensions as follows

\begin{equation}
\rho(\overrightarrow{n})=\frac{1}{2}\mathbf{1+\sum_{i=1}^{n^{2}-1}}c_{i}\mathbf{g}_{i}\end{equation}

with $\mathbf{g}_{i}$ being the $SU(n)$ generators, with the assumption
that if $\mathbf{\sum_{i=1}^{n^{2}-1}}\left|c_{i}\right|^{2}=\frac{1}{4}$
then the density matrix represents a pure state. 

Utilizing QLib's parametrization capabilities, we shall generate a
large number of random pure states, separable states and general density
matrices and plot the $2d$ projections of the resulting Bloch {}``hyper-sphere'',
i.e. scatter plots of two components of $\overrightarrow{c}$. It
is evident from figure 3 that no such trivial generalization is possible,
and that the geometry of the problem is far more complex that can
be naively guessed.

\begin{figure}
\includegraphics[scale=0.9]{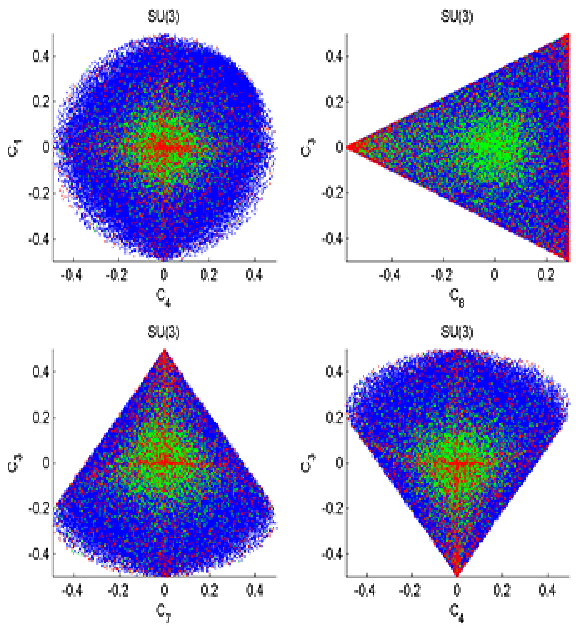}\smallskip{}
\includegraphics[scale=0.9]{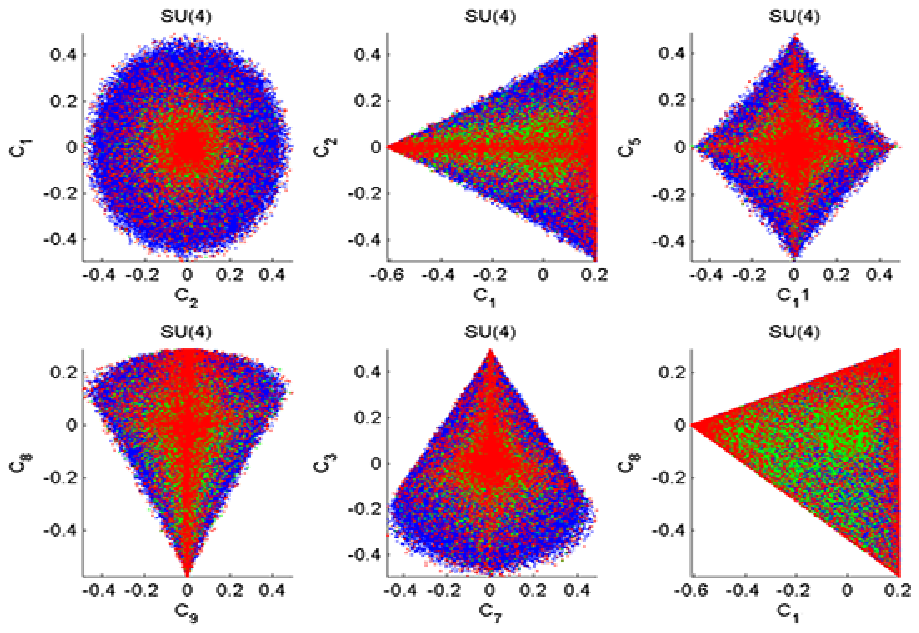}

\caption{(color on-line) $2d$ projections of the SU(3) and SU(4) Bloch {}``hyper-spheres''.
Blue dots indicate general density matrices. Green are separable states
and red dots indicate pure states.}

\end{figure}

\subsection{Additivity of entanglement and entropy measures}

Another simple use of QLib is to experimentally test the additivity
of entropy and entanglement measures\begin{equation}
E(\rho_{1}\otimes\rho_{2})\overset{?}{=}E(\rho_{1})+E(\rho_{2})\end{equation}

By randomly generating multiple $\rho_{1}$-s and $\rho_{2}$-s and
checking the additivity attribute for each, we can form a reliable
hypothesis regarding the behavior of the measure in question. Moreover,
by extremizing $E(\rho_{1}\otimes\rho_{2})-E(\rho_{1})-E(\rho_{2})$
over all possible $\rho_{1}$,$\rho_{2}$ one may reach an even more
well-founded conclusion. Of particular interest is the relative entanglement
measure \protect\cite{rel-ent vedral}

\begin{equation}
E_{R}=\underset{{\scriptstyle \sigma\in SEP}}{\textrm{inf}}tr\,\rho\left(\textrm{log}\rho-\textrm{log}\sigma\right)\end{equation}
which is a generalization of the classical relative entropy\begin{equation}
S(p|q)=tr\, p(\log p-\log q)\end{equation}

It is known that $E_{R}$ is non-additive \protect\cite{rel-ent volbrecht werner}. 

To compute $E_{R}$, one must be able to compute $\underset{{\scriptstyle \sigma\in SEP}}{\textrm{inf}}$,
which in turn requires parametrization of the separable space. In
QLib this is achieved using the observation of P. Horodecki \protect\cite{sep param horodecki},
that the separable space is convex, and thus each point within is
constructable as a linear interpolation of a finite number of extremal
points of that space, as per the Caratheodory theorem. Therefore,
to parametrize all separable density matrices of dimension $d$, one
may parametrize $d^{2}$ separable pure states of the same dimensionality
$|\phi_{i}\rangle$ and a classic probability distribution $\{ p_{i}\}$
to specify the mixing, resulting in the parametrization 

\begin{equation}
\rho=\sum_{i=1}^{d^{2}}p_{i}|\phi_{i}\rangle\langle\phi_{i}|.\end{equation}

\smallskip{}

The numerical study of $E_{R}$ additivity clearly indicate that the
relative entanglement is super-additive, i.e.

\begin{equation}
\textrm{hypothesis:}\,\,\,\,\,\,\,\, E_{R}(\rho_{1}\otimes\rho_{2})>E_{R}(\rho_{1})+E_{R}(\rho_{2})\end{equation}
\smallskip{}

\section{Looking forward - a community effort}

QLib is distributed as free software. The word \char`\"{}free\char`\"{}
does not only refer to price; primarily it refers to freedom: You
may run the program, for any purpose, study how it works, adapt it
to your needs, redistribute copies and improve the program. 

\smallskip{}
It is our hope is that QLib will evolve into a group effort, maintained,
nurtured and grown by the Quantum Information community, for the benefit
of us all. For that purpose, we have licensed QLib under the GPL,
or GNU Public License, which sets-up both the freedom to use the software,
and the requirement that any enhancements made to QLib be released
back to the community. Code which uses QLib, but is not part of it,
may, of course, remain private. For more information regarding these
issues, see the licensing section of the QLib website. 

Several tools are available on the website to facilitate joint development
of future versions: Forums, a bug tracking system, a feature request
form and a mailing list. \smallskip{}

The direction future QLib development will take shall be determined
by you, its users.

\smallskip{}

\begin{acknowledgments}
We thank B. Reznik for allowing the freedom to explore uncharted waters. 

This work was supported by the Israeli Science Foundation (Grants
784-06 and 990-06).
\end{acknowledgments}

\end{document}